\documentclass[notitlepage,nofootinbib, superscriptaddress, 12pt, a4paper, twocolumn]{revtex4-1}

\usepackage{graphicx}
\usepackage{dcolumn}
\usepackage{bm}
\usepackage{xspace}
\usepackage{amsmath,amsfonts,amssymb}
\usepackage[colorlinks]{hyperref}
\usepackage{units,nicefrac}
\usepackage{color}
\usepackage[normalem]{ulem}

\newcommand{\snnt}[1]{\ensuremath{\sqrt{s_{NN}} = #1 \text{ TeV}}\xspace}
\newcommand{\snng}[1]{\ensuremath{\sqrt{s_{NN}} = #1 \text{ GeV}}\xspace}

\newcommand{\gevc}[1]{\ensuremath{#1 \text{ GeV/$c$}}\xspace}

\newcommand{\rhoc}{\ensuremath{\rho}\xspace}

\newcommand{\pizero}{\ensuremath{\pi^{0}}\xspace}

\newcommand{\pp}{p-p\xspace}
\newcommand{\nuclnucl}{A-A\xspace}

\newcommand{\pT}{\ensuremath{p_{\text{T}}}\xspace}
\newcommand{\pTinitial}{\ensuremath{p_{\text{T,i}}}\xspace}
\newcommand{\pTfinal}{\ensuremath{p_{\text{T,m}}}\xspace}
\newcommand{\dpT}{\ensuremath{\delta_{\pT}}\xspace}
\newcommand{\DpT}{\ensuremath{\Delta \pT}\xspace}
\newcommand{\RAA}{\ensuremath{R_{\text{AA}}}\xspace}
\newcommand{\RAAin}{\ensuremath{R_{\text{AA, in}}}\xspace}
\newcommand{\RAAout}{\ensuremath{R_{\text{AA, out}}}\xspace}
\newcommand{\vtwo}{\ensuremath{v_{2}}\xspace}
\newcommand{\vthree}{\ensuremath{v_{3}}\xspace}
\newcommand{\dd}{\text{d}}

\newcommand{\Lin}{L_\text{in}}
\newcommand{\Lout}{L_\text{out}}

\newcommand{\onehalf}{{\nicefrac{1}{2}}}
\newcommand{\threequarters}{{\nicefrac{3}{4}}}
\newcommand{\threeeights}{{\nicefrac{3}{8}}}

\begin{document}

\title{Universal scaling dependence of QCD energy loss from data driven studies}

\author{P. Christiansen} 
\affiliation{Division of Particle Physics, Lund University, Sweden}

\author{K. Tywoniuk}
\affiliation{Departament d'Estructura i Constituents de la Materia and Institut de Ci{\`e}ncies del Cosmos, Universitat de Barcelona, Mart{\'i} i Franqu{\'e}s 1, ES-80 028 Barcelona, Spain}

\author{V. Vislavicius}
\affiliation{Division of Particle Physics, Lund University, Sweden}

\date{\today}

\begin{abstract}
  In this paper we study the energy loss of jets in the QGP via the nuclear modification factor \RAA for unidentified particles at high \pT ($\gtrsim \gevc{10}$) in and out of the reaction
  plane of the collision. We
  argue that at such a high \pT there are no genuine flow effects and, assuming
  that the energy loss is only sensitive to initial characteristics such as the density and geometry, find that
  \RAA depends linearly on the (RMS) length extracted from Glauber
  simulations. Furthermore we observe that for different centrality classes
  the density dependence of the energy loss enters as the square root of the
  charged particle multiplicity normalized to the initial overlap
  area. The energy loss extracted for RHIC and LHC data from the \RAA is found to exhibit a universal behavior.
\end{abstract}

\maketitle

\section{Introduction}
\label{sec:intro}

One of the most stunning results from the heavy ion programs at RHIC and LHC is the quenching of jets and single-inclusive hadron spectra~\cite{Adcox:2001jp,Aamodt:2010jd,Aad:2010bu}. Being perturbative probes for which we can calculate the vacuum baseline to high precision, jets are potentially excellent probes of the medium created in heavy ion collisions. Modifications, arising due to interactions with the hot and dense matter, are indeed expected to arise at timescales comparable to the lifetime of the medium are typically characterized in terms of elastic and radiative energy losses \cite{Gyulassy:1990ye,Wang:1991xy}, for recent reviews see, e.g., \cite{d'Enterria:2009am,Majumder:2010qh,Mehtar-Tani:2013pia}. Presently our theoretical control of the jet fragmentation is however limited. In particular, the importance of modifications of the jet sub-structures due to the transverse medium resolution was only recently pointed out \cite{CasalderreySolana:2012ef}. Recent results from the LHC on the suppression of single-inclusive hadrons and jets are in this context challenging to reconcile with the corresponding observations at RHIC \cite{Horowitz:2011gd} and call for the refinement of present theoretical tools. Furthermore, at RHIC it is challenging to reconcile both the data on the nuclear modification factor, \RAA, and azimuthal flow, characterized by \vtwo, at high \pT within models based on radiative QCD mechanisms \cite{Adare:2010sp,Adare:2012wg}. We take this uncertainty at the theoretical level as an opportunity to make a data driven study that we present here. Similar studies have also been carried out previously in~\cite{Lacey:2009ps,Lacey:2012bg,Lacey:2012kb}, see also \cite{Betz:2011tu} for more theoretically driven studies, and we will return to how they differ from the present work in Section~\ref{sec:discussion}.

One of the challenges of modeling the energy loss is that the medium created in heavy ion collisions behaves as a perfect liquid. There are at least 2 major issues. First of all, both the geometry and the dynamical expansion of the medium introduce a complication for the clean extraction of the transport properties of the medium. The longitudinal expansion of the medium causes the energy density to decrease quickly with time (as the inverse of the proper time in the Bjorken model~\cite{Bjorken:1982qr}), and this could clearly affect the path length dependence of the energy loss. Furthermore, the dynamics of the medium in the transverse plane signifies that in non-central collisions there is an asymmetric expansion of the medium, where the expansion in the reaction plane is larger than out-of-plane. To first order the latter effect is supposed to be negligible, but various studies have documented significant effects~\cite{Renk:2010qx}.  Since we wish to pursue a data driven study, these effects cannot be handled without recourse to modeling and so we will focus on characterizing the energy loss in terms of initial state observables. It is quite remarkable that this seems to work very well.

Secondly, the convincing signals of collective behavior in \nuclnucl collisions hint at the existence of a strongly coupled system. This, in turn, challenges the paradigm of using perturbative methods to calculate the relevant degrees of freedom for the jet-medium interactions. Our present study avoids these conceptual difficulties.

One could worry that the measured \RAA in and out of the reaction plane is
significantly affected by flow. Let us try to argue here that for $\pT >
\gevc{8}$ this is in our opinion not very likely. Flow is typically
characterized by introducing a mass dependence. Both measurements of
\vtwo~\cite{Abelev:2012di} and the \RAA~\cite{OrtizVelasquez:2012te} have
shown that for $\pT > \gevc{8}$ there is little or no difference between
results for pions and protons. The triangular flow, characterized by the
coefficient \vthree, also seems to disappear in this \pT
region~\cite{Abelev:2012di}. As the baryon to meson ratios are rather similar
from RHIC energy (\snng{200}) to LHC energies
(\snnt{2.76})~\cite{Abelev:2013xaa} this indicates that also for RHIC energies
we need to have data for $\pT > \gevc{8}$ to eliminate flow effects.  In our
opinion, this allows us safely to assume that both \RAA and \vtwo at high \pT
are dominated by energy loss. The effect of residual flow would be an
underestimate (overestimate) of the quenching contribution in (out) of
plane. There are no indications for such an effect in Fig.~\ref{fig:scaling}.

At high energies the energy loss of a colored parton going through a colored medium is expected
to be dominantly radiative. Na\"ively one expects that radiative QCD energy loss \cite{Baier:1998kq,Gyulassy:2000er,Wiedemann:2000za,Qiu:1990xxa,Arnold:2002ja} increases quadratic with path length, since this follows from the stimulated emission probability of a single \emph{hard} gluon \cite{Baier:1996kr,Zakharov:1997uu}. These emissions are however rare and one should also account for multiple \emph{soft} emissions. This changes the path length dependence of the characteristic \pT shift of the medium-modified spectra so it becomes linear \cite{Baier:2001yt}, see Appendix~\ref{sec:app:radiative}. A crucial point of this paper is that the existing data allows to disentangle more than simply the path length dependence of the suppression. As we will show, this information has to be supplemented by including a dependence on the energy density. While our results will rely on simple estimates of both of these quantities, for details see Section~\ref{sec:setup}, the agreement with the data at two, widely separated energies of RHIC and LHC represent a strong argument for the consistency of the interpretation of energy loss in ultrarelativistic heavy ion collisions. Finally we note that there is a significant \pT dependence of the \RAA. We shall ignore this \pT dependence in our quantitative studies and focus on a common \pT region of $\pT \approx \gevc{10}$ for LHC and RHIC. The scaling plots we show, in particular Fig.~\ref{fig:pt_ratio}, does however indicate that the scaling relations we find are also valid at higher \pT. 

The outline of the paper is as follows. In Section~\ref{sec:setup} we describe the data driven set-up and the method for extracting the energy loss (or \pT shift) from the \nuclnucl spectra. We go on to present the obtained results and discuss them in Sections~\ref{sec:results} and \ref{sec:discussion}, respectively. Finally, our conclusions are summarized in Section~\ref{sec:conclusions}. 

\section{Data driven set-up}
\label{sec:setup}

\begin{figure}[tbp]
  \begin{center}
    \includegraphics[keepaspectratio, width=0.49\columnwidth]{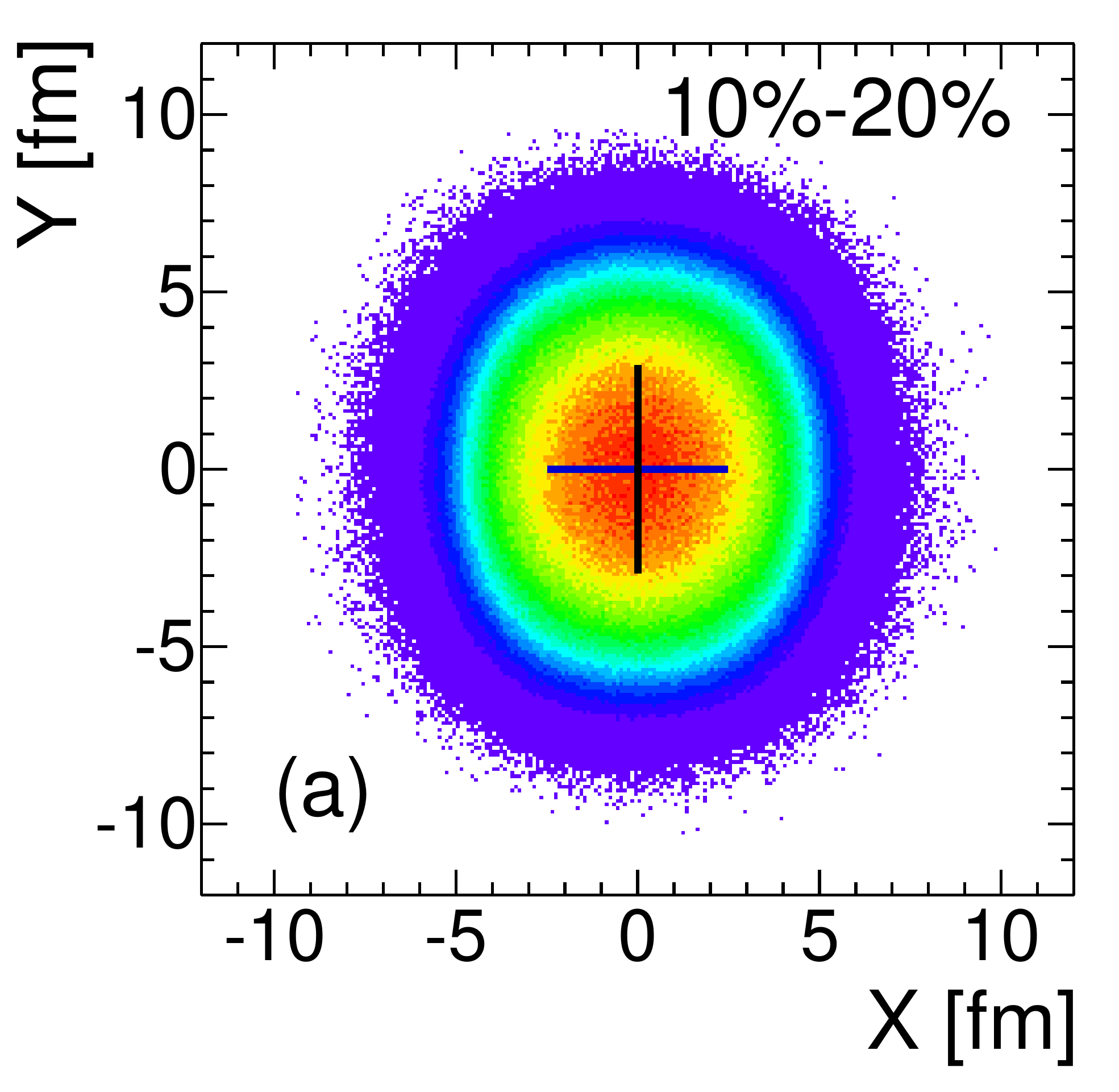}
    \includegraphics[keepaspectratio, width=0.49\columnwidth]{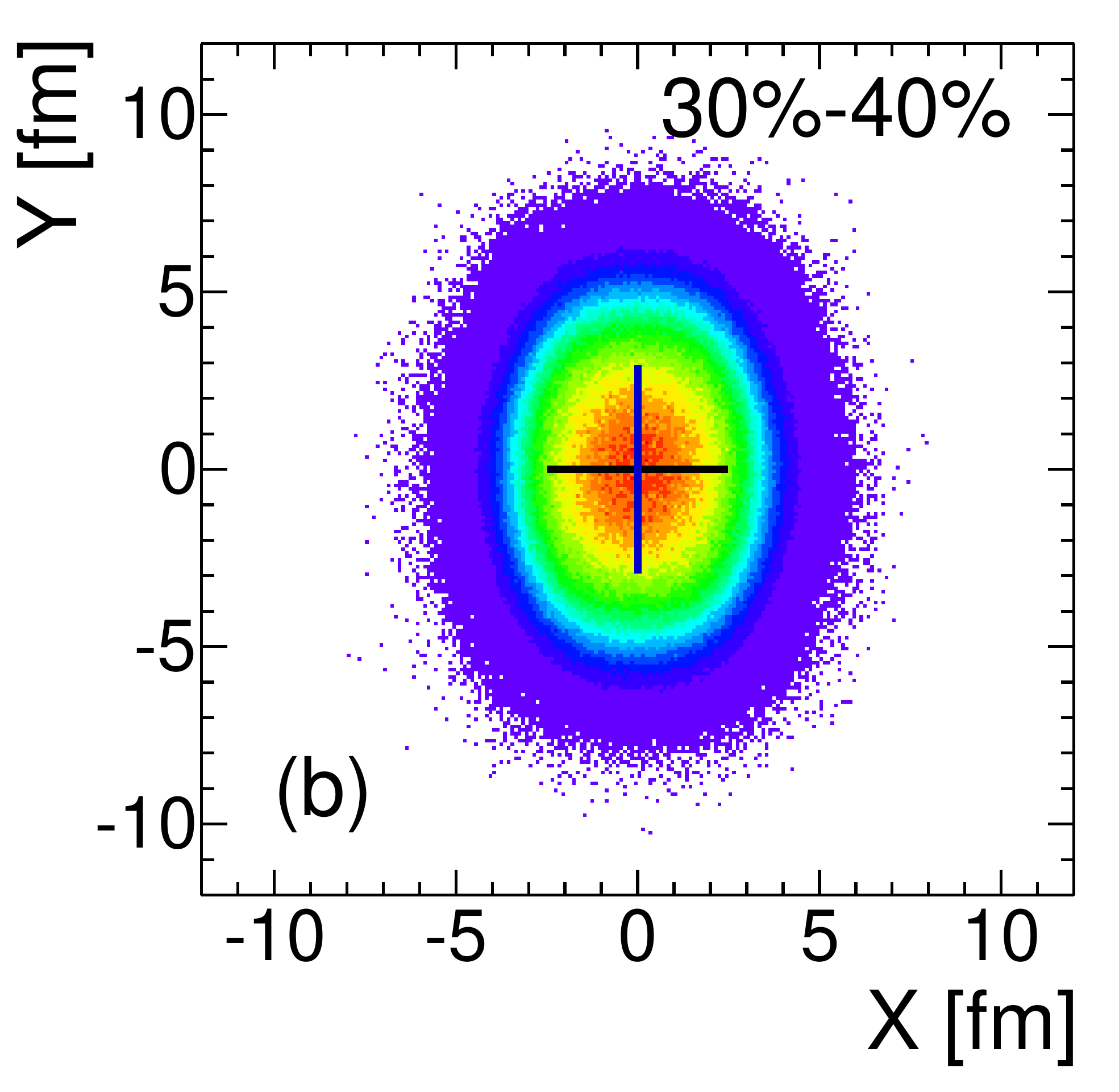}
    \includegraphics[keepaspectratio, width=0.49\columnwidth]{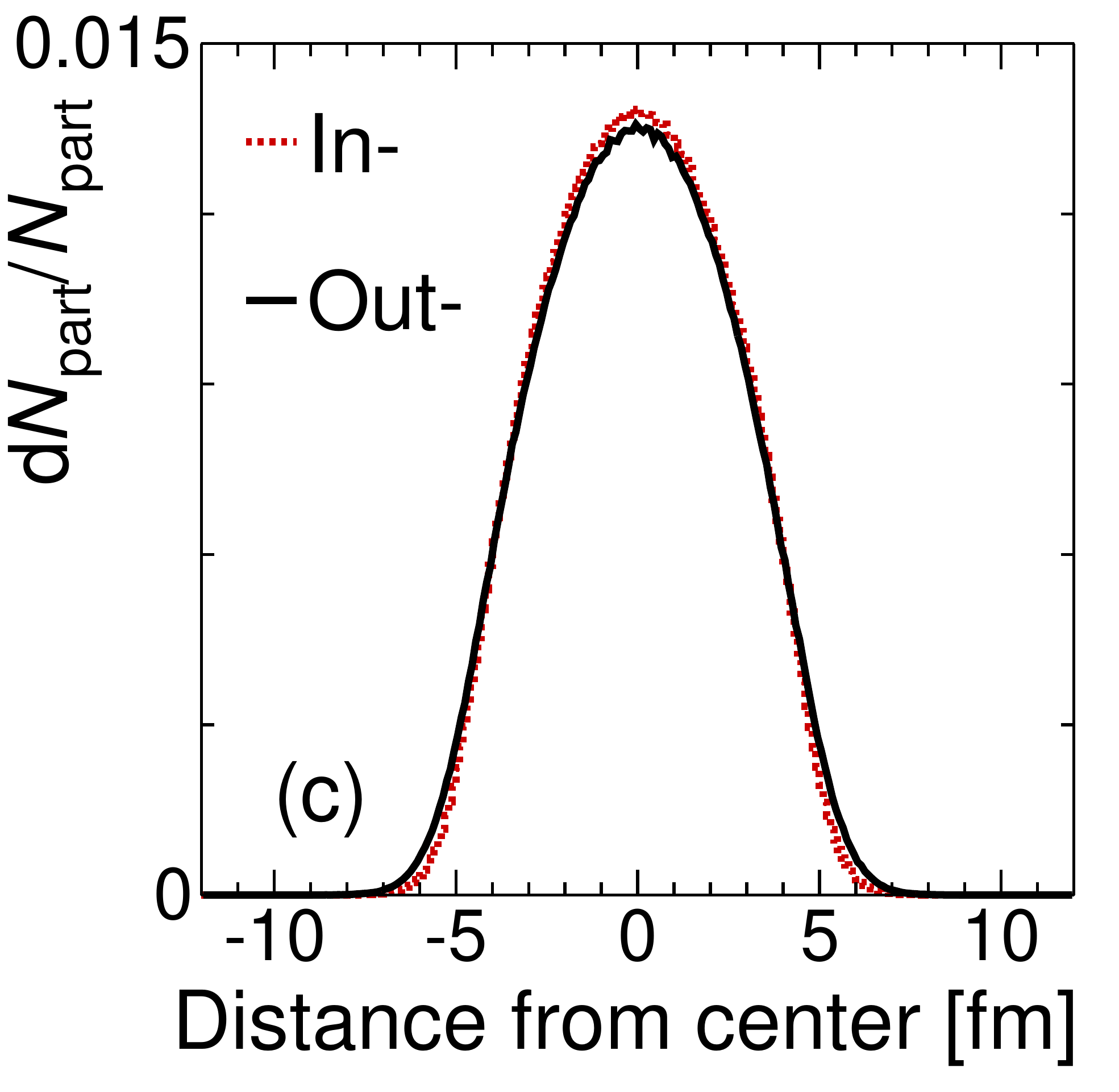}
    \includegraphics[keepaspectratio, width=0.49\columnwidth]{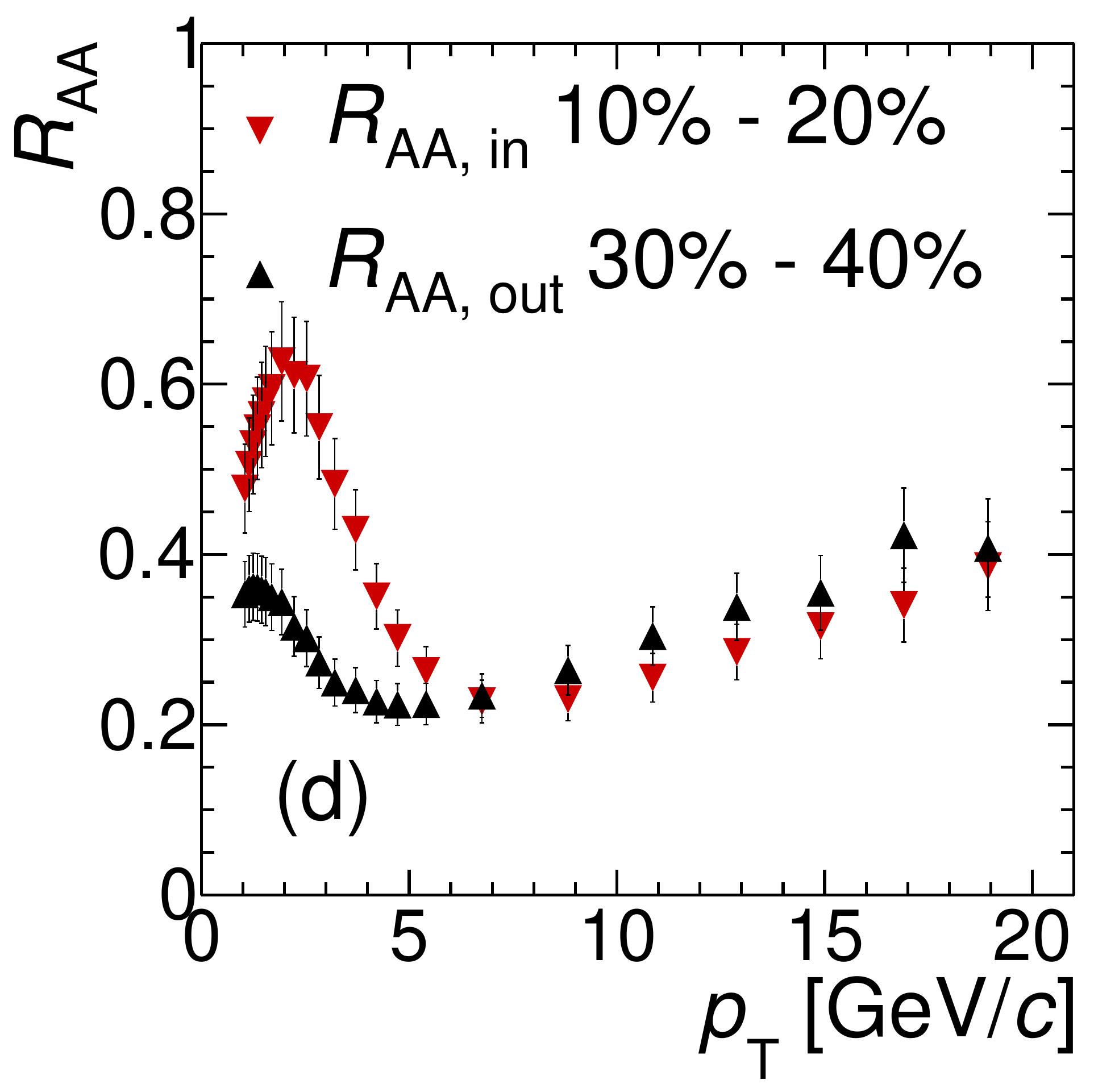}
  \end{center}
  \caption{(Color online) The extracted participant distribution for two Glauber samples at \snnt{2.76} (10\%-20~\% centrality (a) and 30\%-40~\% centrality (b)) rotated so that the reaction plane coincides with the $x$-axis. The in-plane RMS of the former equals approximately the out-of-plane of the latter. Comparison of the participant distributions and the \RAA for the two cases are shown in (c) and (d), respectively.}
  \label{fig:L_example}
\end{figure}

Figure~\ref{fig:L_example} illustrates the idea behind the studies presented here. Based on Glauber simulations of the participant distribution, two centrality classes are selected where we can relate some properties in- vs. out-of-plane. In our case the selection was done on the characteristic length, which we define as the root-mean-squared (RMS) of the distribution. We can then compare the in- and out-of-plane \RAA data for different centrality classes where these properties will agree. In Fig.~\ref{fig:L_example} we have chosen two centrality classes, 10-20~\% and 30-40~\%, where the in-plane RMS width of the former distribution, denoted $\Lin$, is approximately equal to the out-of-plane RMS, denoted $\Lout$, of the latter, see the top panels. In the lower-left panel of Fig.~\ref{fig:L_example} we also demonstrate that the participant distribution is quite similar in the two cases. 

An important motivation behind such a simplified event selection is the fact that in central collisions we expect the distribution of hard scatterings (binary collisions) to be more narrowly distributed around the origin. In that way the path length of the two samples should on the average be quite similar, but we note most importantly that the density is quite different. Moreover, the transverse expansion could be much more significant in-plane than out-of-plane and could spoil the comparison. In our studies we find that the latter effect can be neglected and this is in fact also, as mentioned above, what one would expect to first order from theoretical arguments.

Once we have fixed the characteristic length to be similar, it remains to include the effect of the difference in energy density. As it is clearly seen in the lower-right panel of Fig.~\ref{fig:L_example}, comparing the \RAA for our example cases for which the path lengths were equal in- and out-of-plane does not result in the same amount of suppression.
The overlapping participant distributions are reasonably described by two-dimensional Gaussian distributions, see lower-left panel of Fig.~\ref{fig:L_example}, and so we assign an area as $A \approx 4\pi \Lin \Lout$. Then we assume that the characteristic energy density $\rhoc$ of the sample is given by
\begin{equation}
\label{eq:BjorkenDensity}
\rhoc = K \frac{dN/d\eta}{4\pi \Lin \Lout},
\end{equation}
where $K$ is a constant that is assumed to depend little on centrality and
collision energy. In the following we always set $K=1$~GeV/fm such to make \rhoc
have the units GeV/fm$^3$. As this density is not normalized in a meaningful
way (because of the data driven nature of this study) we will in the following use arbitrary units (arb. units) in the plots. The pseudorapidity distribution, $dN/d\eta$, have been taken
from~\cite{Aamodt:2010cz}. In Sec.~\ref{sec:discussion} where we introduce
theoretical estimates for comparison we will discuss how one can normalize
this properly to extract meaningful physics parameters. The definition of
\rhoc is inspired by Bjorken's energy density estimate and the observation
that the mean transverse energy per produced particle does not change
violently as a function of centrality or collisions
energy~\cite{Chatrchyan:2012mb}.

The LHC data on charged particle \RAA and \vtwo used in this publication have
been taken from~\cite{Abelev:2012hxa,ATLAS:2011ah}. CMS has published similar
data~\cite{CMS:2012aa,Chatrchyan:2012xq} but with coarser segmentation in
centrality and \pT, while ALICE \vtwo measurements does not cover centralities
above 50~\%~\cite{Abelev:2012di}. The \RAA in- and out-of-plane used in our
data driven analysis has been obtained as $\RAAin = \RAA (1+2 \vtwo)$ and
$\RAAout = \RAA (1-2 \vtwo)$, respectively. The \pT bins for the \RAA and
\vtwo results do not match perfectly but the closest \pT points have been used
and as both the \RAA and \vtwo are only rather moderately varying at high \pT
we consider this a negligible effect. The error bars shown in the figures for
\RAAin and \RAAout always include the full statistical and systematic
uncertainties added in quadrature from both the \RAA and \vtwo. Normalization
errors for \RAA have been ignored as they are expected to be directly
correlated across centralities (and to some degree also across beam
energies). When $\RAAin$ and $\RAAout$ is compared we assume in our
interpretation that the relative systematic error is smaller than shown. For
the \RAA one expects e.g. the efficiency and corrections to have similar
systematic errors and so there it seems a common shift of $\RAAin$ and
$\RAAout$ is expected. On the other hand for \vtwo a systematic shift would
tend to shift $\RAAin$ and $\RAAout$ in opposite directions. A better
understanding of this aspect can only be obtained by the experiments.

To extract information beyond merely the level of suppression of the spectra,
we would like to study the phenomenon of energy loss more
directly~\cite{Adler:2006bw,Adare:2012wg}. To this aim we will assume that the
spectra in \pp and \nuclnucl collisions can be described by a power-law with
a similar exponent and that the difference comes from the
fact that the primordial \pT of the partonic \nuclnucl spectrum has been
shifted to lower values due to energy loss in the medium. Note that the shift
itself could be \pT dependent. Explicitly, the \pT shift is defined as $\DpT
\equiv \pTinitial-\pTfinal$, where $\pTinitial$ is the momentum of the parton
prior to energy loss while $\pTfinal$ is the momentum of the hadron as
measured in the detector. Then, following  a similar method as employed by PHENIX~\cite{Adler:2006bw}, the \pT spectra of particles in a certain centrality class can be compared via
\begin{equation}
\label{eq:SpectraTransformation}
\frac{\dd N_\text{pp}}{\dd \pTinitial}(\pTinitial) = \left|\frac{\dd\pTfinal}{\dd\pTinitial}\right| \RAA(\pTfinal) \frac{\dd N_\text{pp}}{\dd \pTfinal} (\pTfinal),
\end{equation}
where the first term is the Jacobian of the transformation, see Appendix~\ref{sec:app:ptloss} for further details. 
Since we \emph{a priori} cannot predict the dependence of the shift,
we explore two extreme relations between \pTinitial and \pTfinal in
  Eq.~(\ref{eq:SpectraTransformation}): \pT independent absolute and relative
  energy losses (see Appendix~\ref{sec:app:ptloss} for further details). In all figures the central
  value for the \pT loss is the average of the two estimates and the
  systematic uncertainty box shows the actual difference. Here we stress
that the observed scaling patterns are not affected by the resulting
variations in the parameterization of \DpT.

One can find several scaling variables from the orientation-dependent \RAA alone since, e.g., the
squared scaling variable will also align the \RAA. As an additional criterium we will therefore demand that the extracted energy loss is approximately linear in the scaling variable. 

\section{Results}
\label{sec:results}

\begin{figure}[htbp]
  \begin{center}
    \includegraphics[keepaspectratio,width=0.48\columnwidth]{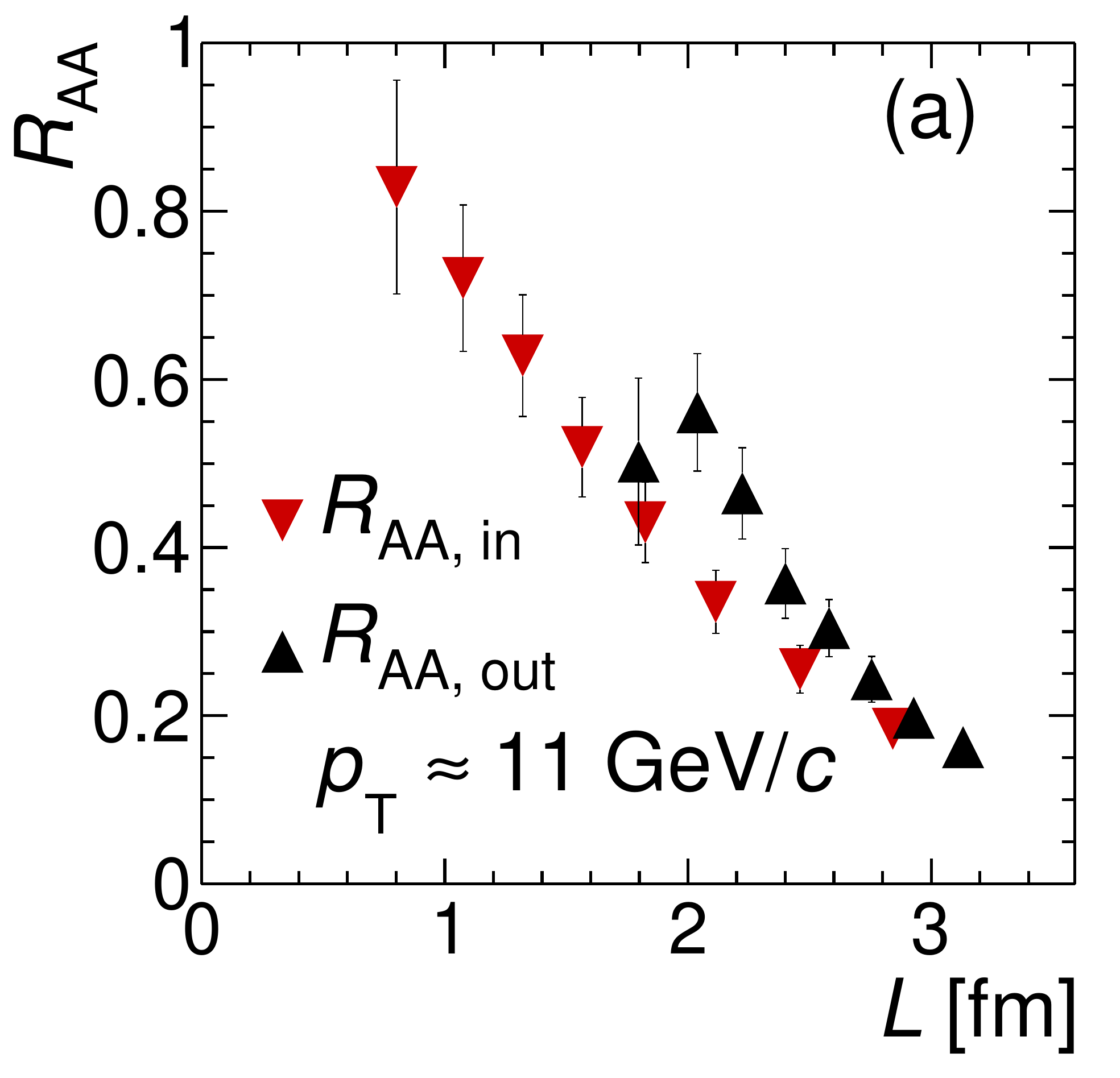}
    \includegraphics[keepaspectratio,width=0.48\columnwidth]{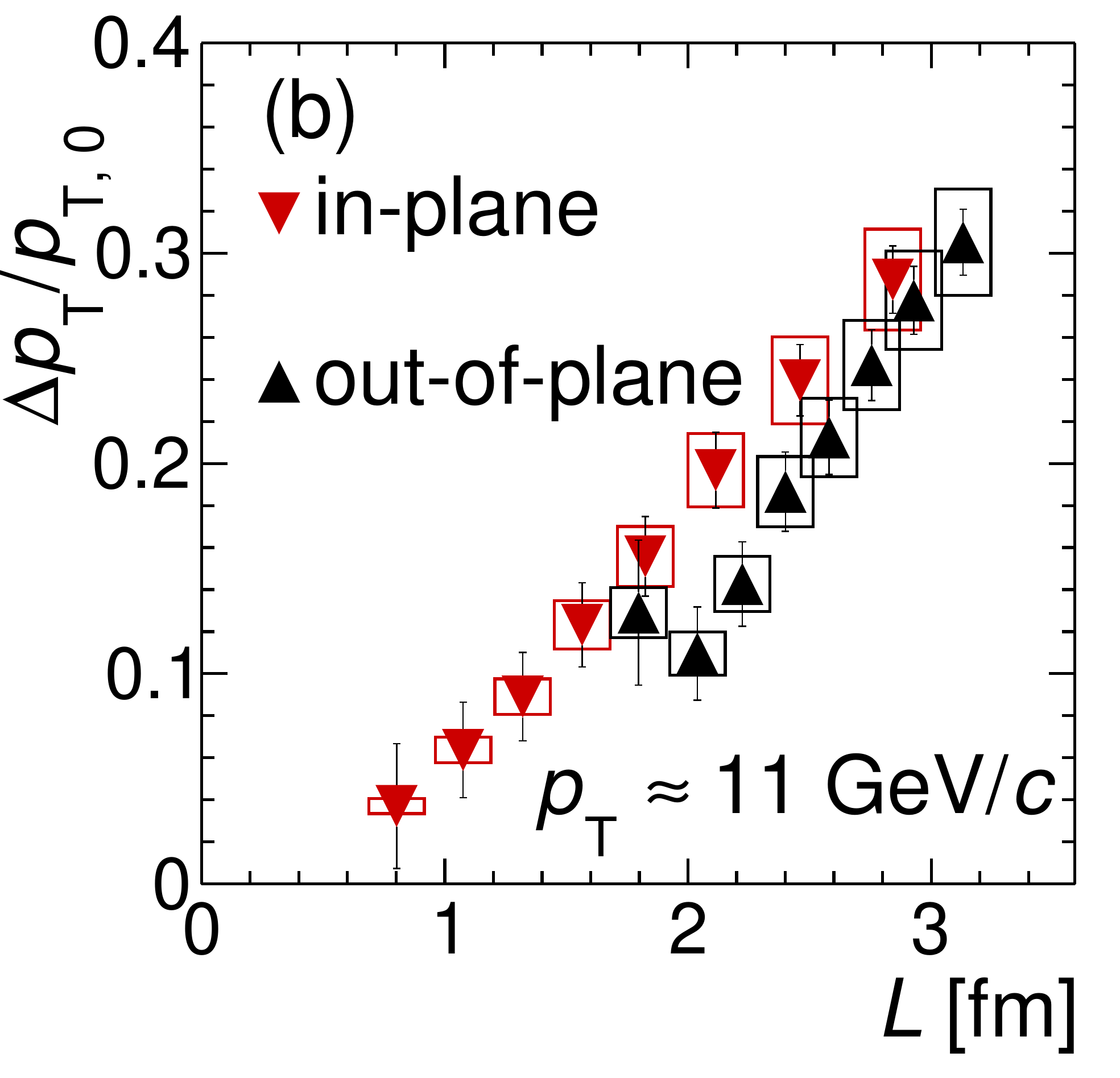}
    \includegraphics[keepaspectratio,width=0.48\columnwidth]{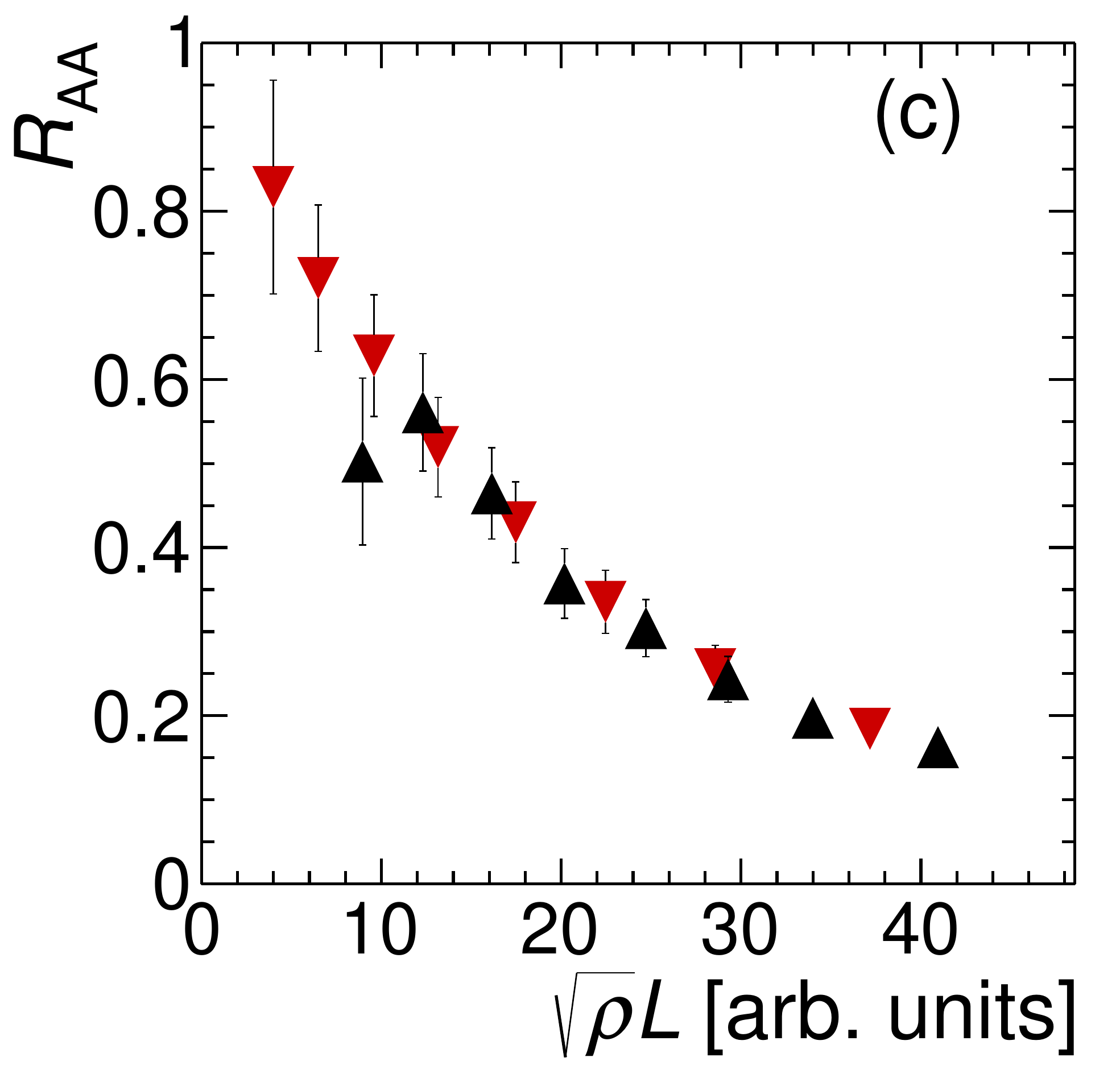}
    \includegraphics[keepaspectratio,width=0.48\columnwidth]{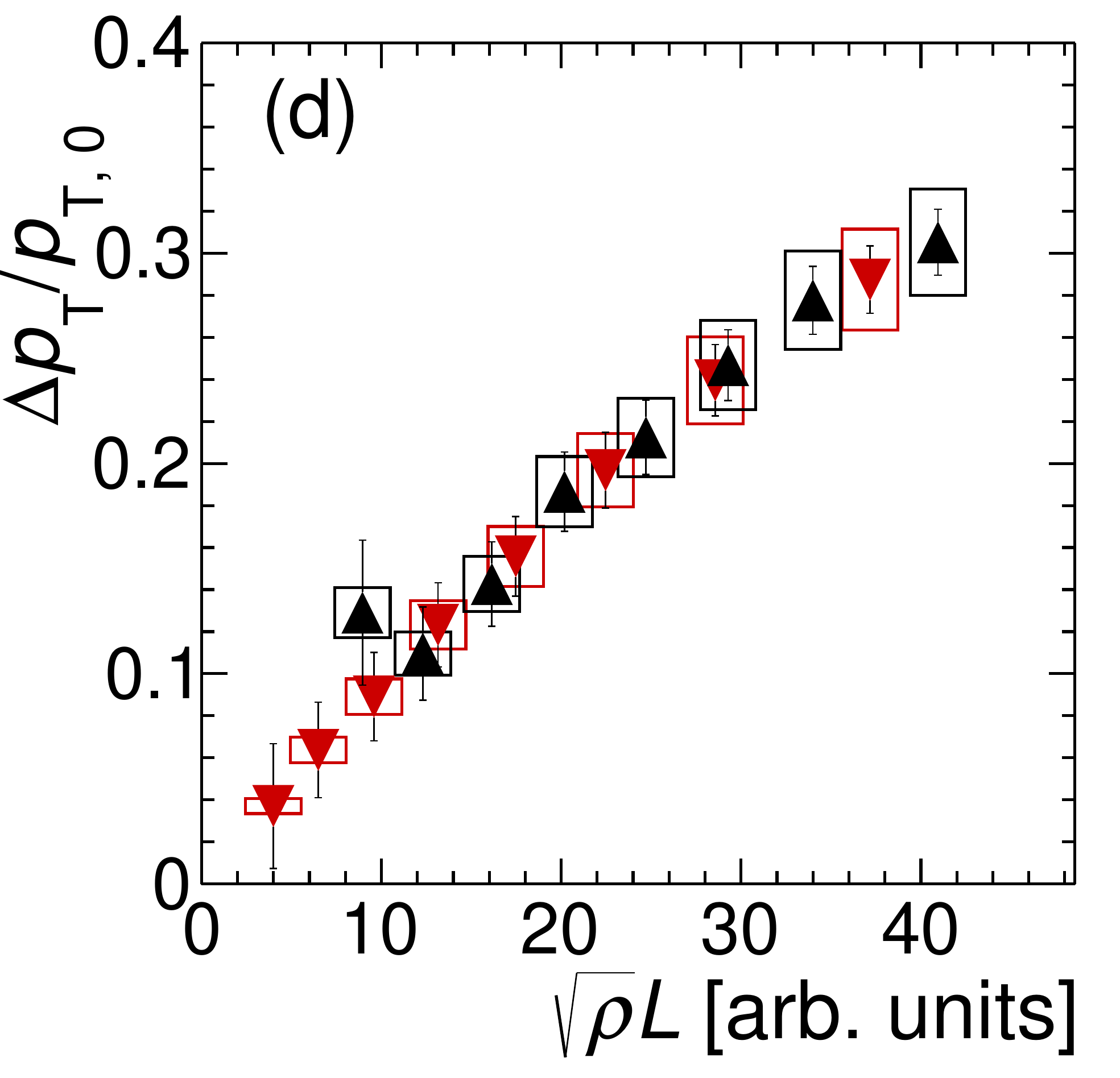}
    \includegraphics[keepaspectratio,width=0.48\columnwidth]{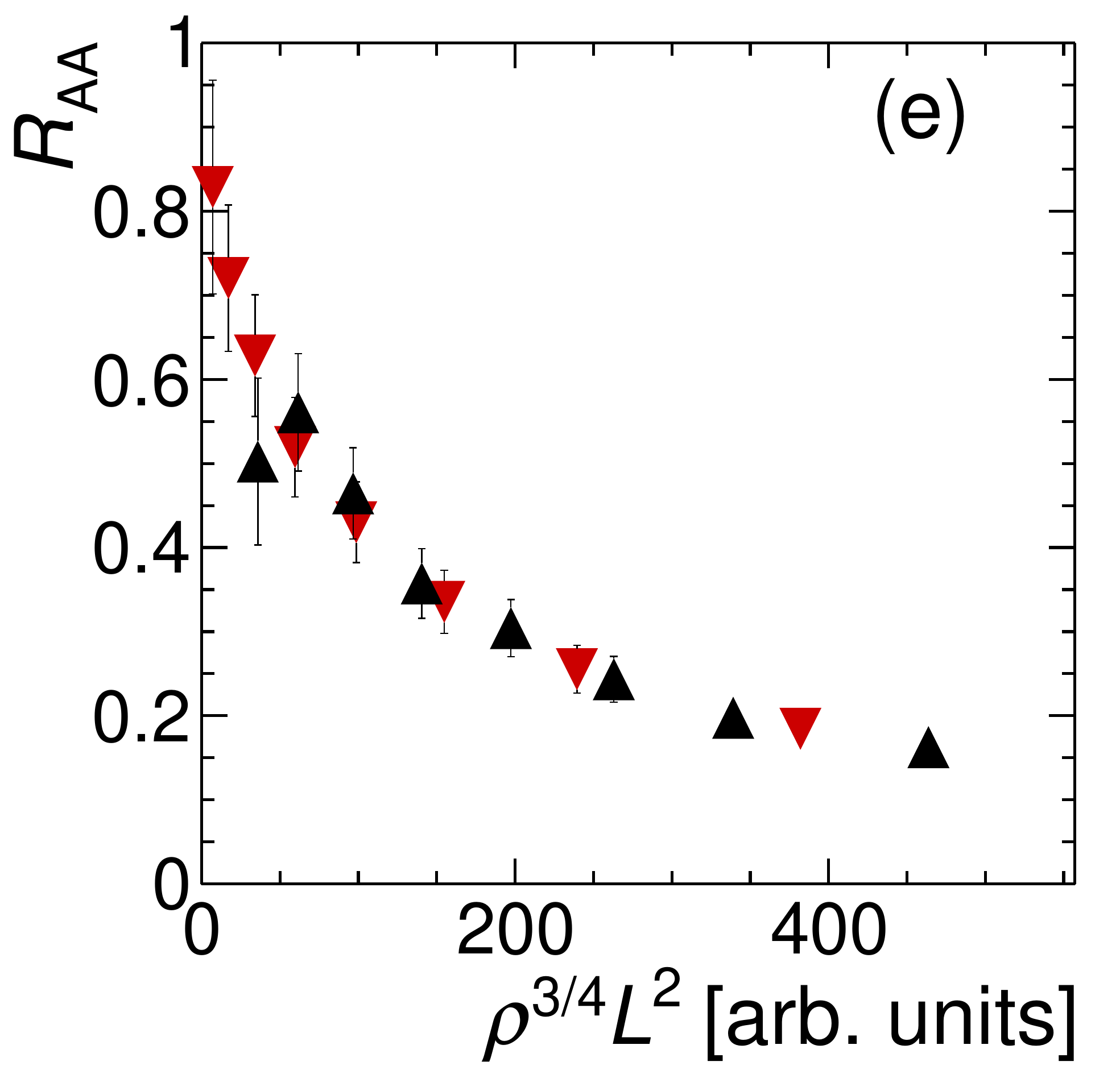}
    \includegraphics[keepaspectratio,width=0.48\columnwidth]{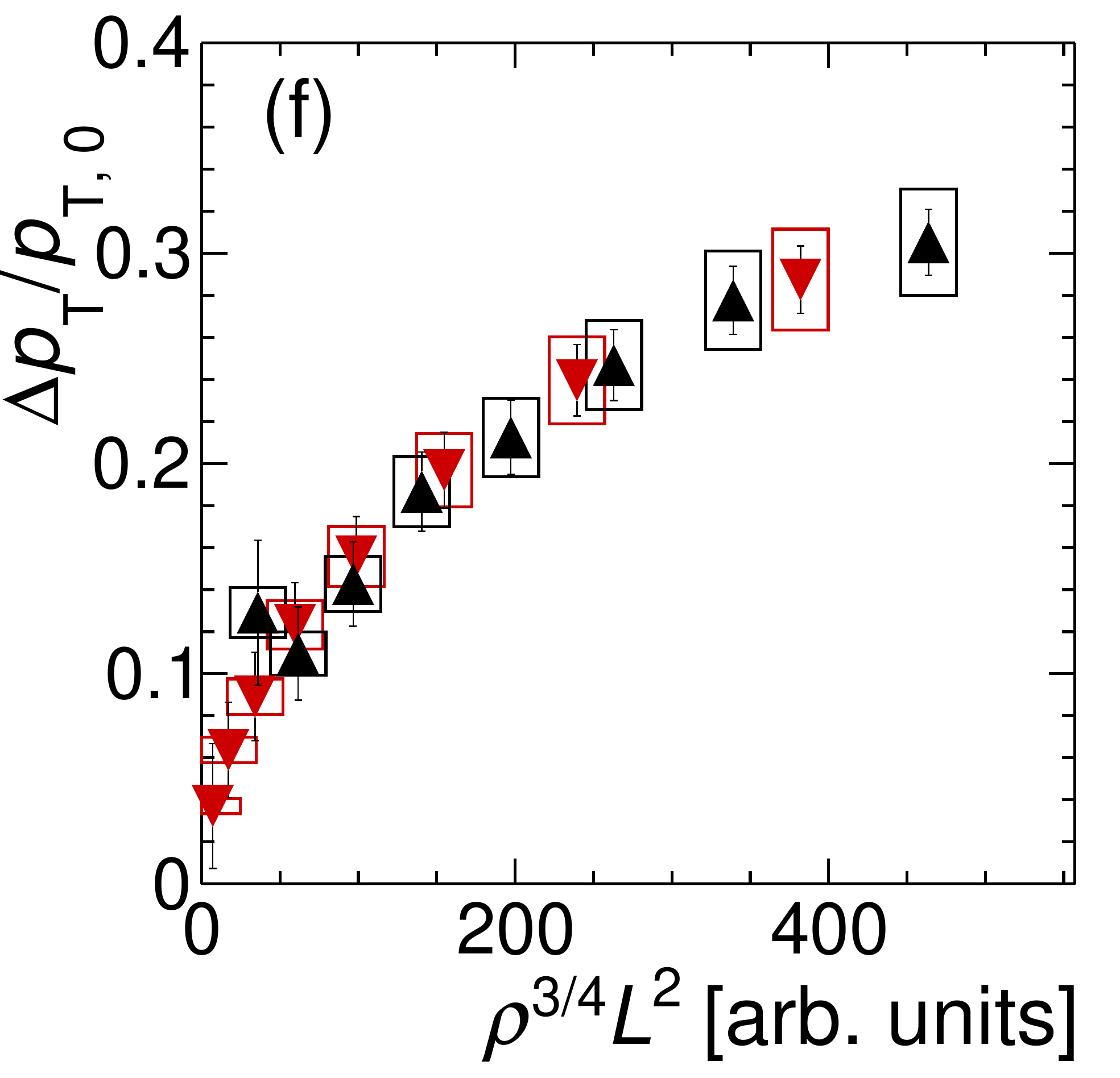}
  \end{center}
  \caption{(Color online) Example of scaling relations for LHC data in arbitrary units. \RAA vs. $L$ (a), $\rhoc^\onehalf L$ (c), $\rhoc^\threequarters L^2$ (e) and extracted energy loss $\DpT/\pT$ vs. the same scaling variables (b, d, f) are shown for $\pT \approx \gevc{13}$. We have included the uncertainty arising from the unknown functional form of $\DpT$ as shaded boxes on the points in the right column, see Appendix~\ref{sec:app:ptloss} for details.}
  \label{fig:scaling}
\end{figure}

Figure~\ref{fig:scaling} shows a summary of the main results from our studies of LHC data. In the left column we plot the \RAA, while in the right one the \pT shift divided by the primordial momentum, $\DpT /\pTinitial$. Both quantities are plotted vs. the respective scaling variable, for which we explore three possibilities: the path length, $L$, in the uppermost row, then $\rhoc^\onehalf L$ in the center and finally $\rhoc^\threequarters L^2$ in the lower column. The motivation behind these choices will be discussed further in Sec.~\ref{sec:discussion}. The plots in the left column illustrate that it is possible to find several scaling variables for the \RAA, but that the energy loss is only approximately linear for the scaling variable in the middle panel. Extrapolating down, it even seems to vanish for $L=0$, as expected. We thus find that all \RAA and \vtwo values for a given \pT
can be described in terms of a linear energy loss relation.
\begin{figure}[htbp]
  \begin{center}
    \includegraphics[keepaspectratio, width=0.49\columnwidth]{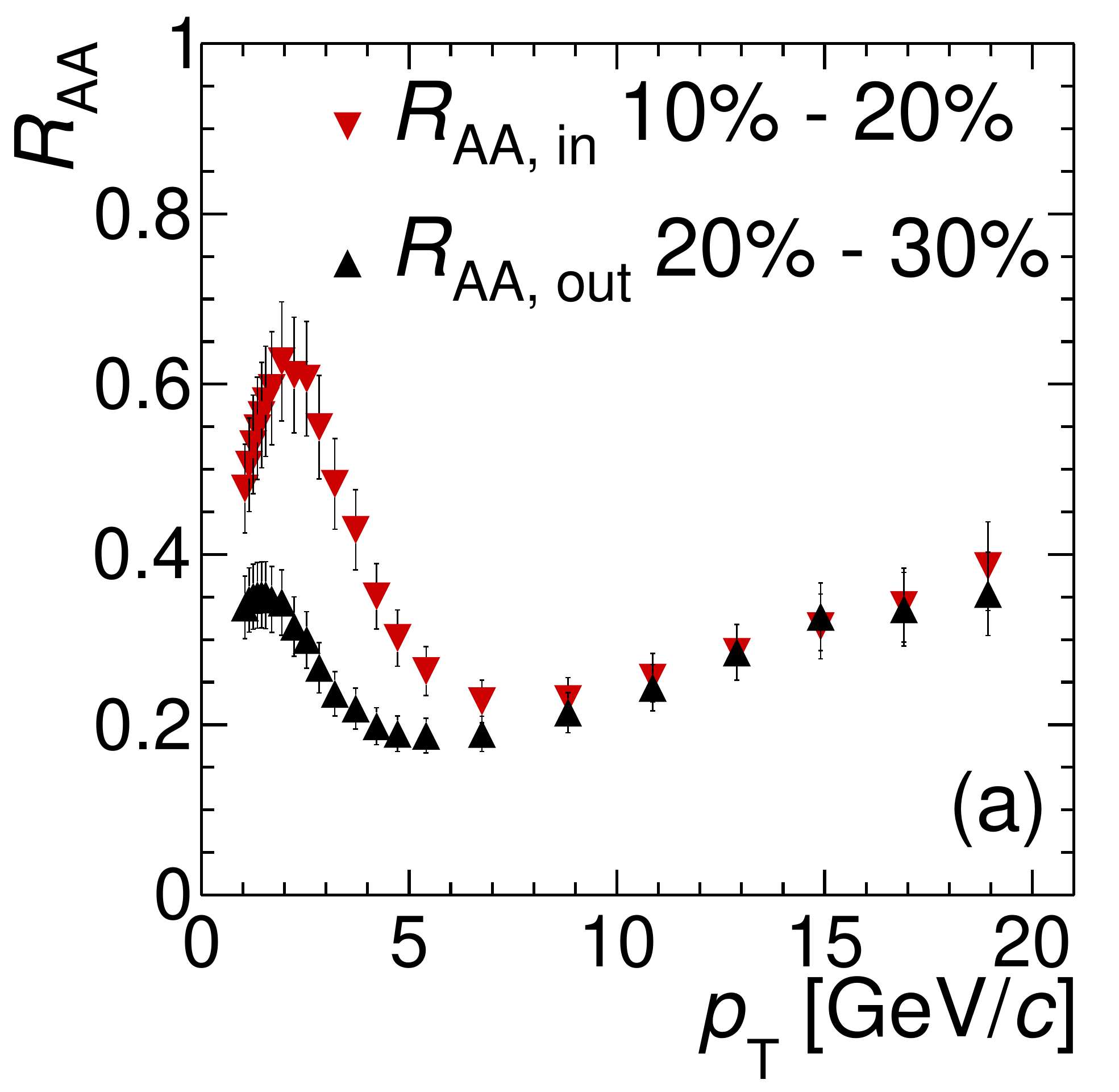}
    \includegraphics[keepaspectratio, width=0.49\columnwidth]{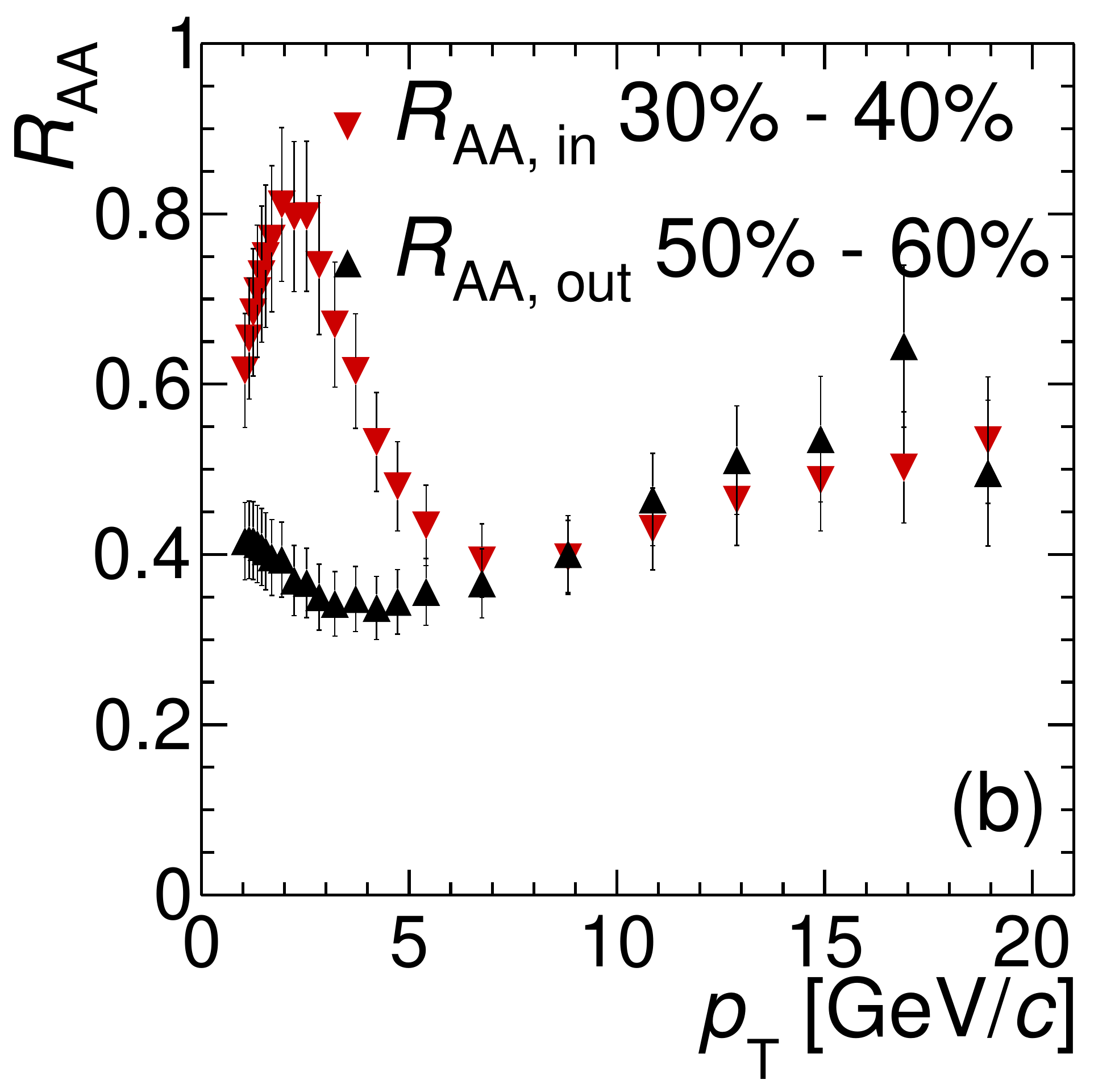}
    \includegraphics[keepaspectratio, width=0.49\columnwidth]{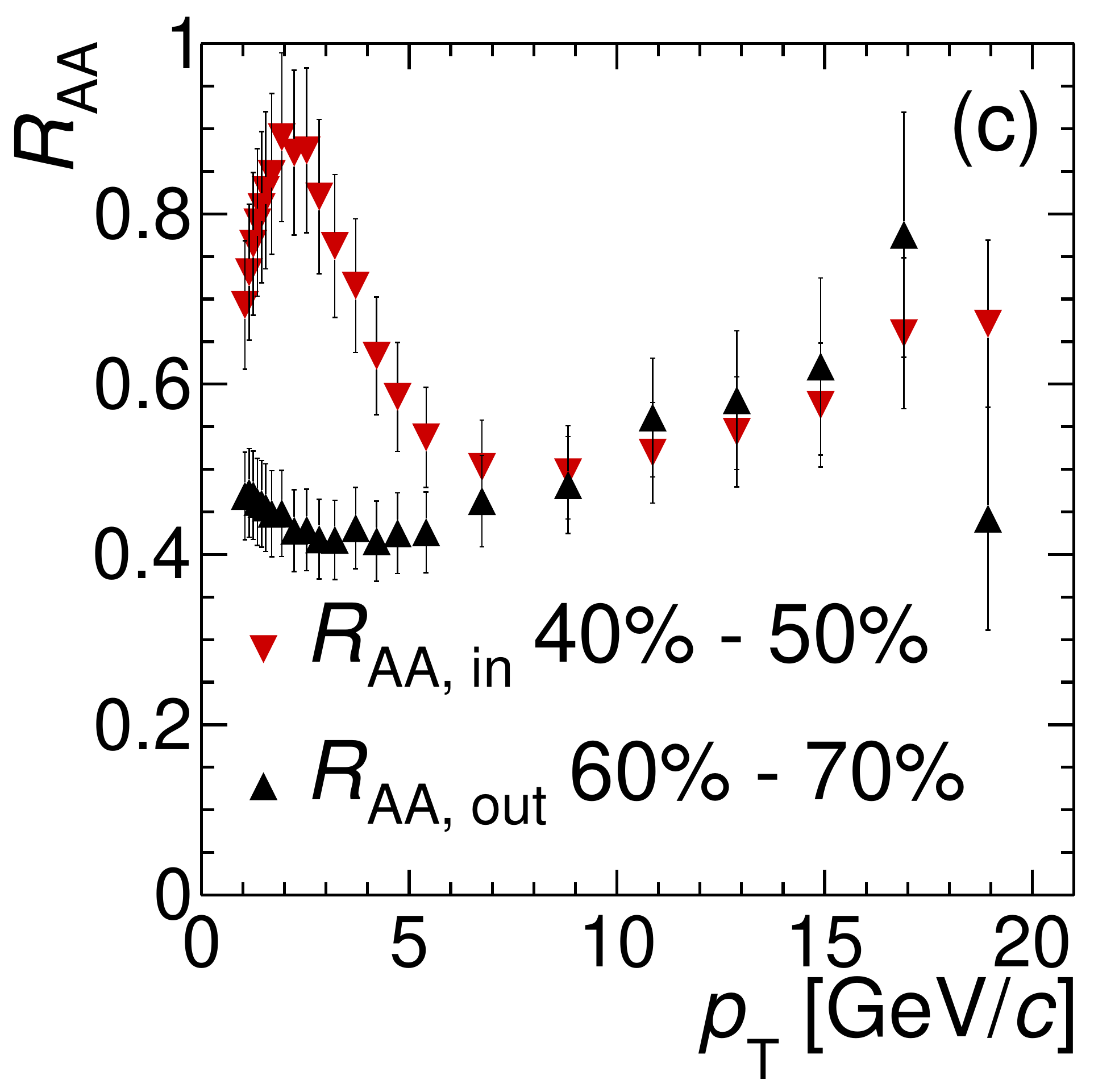}
    \includegraphics[keepaspectratio, width=0.49\columnwidth]{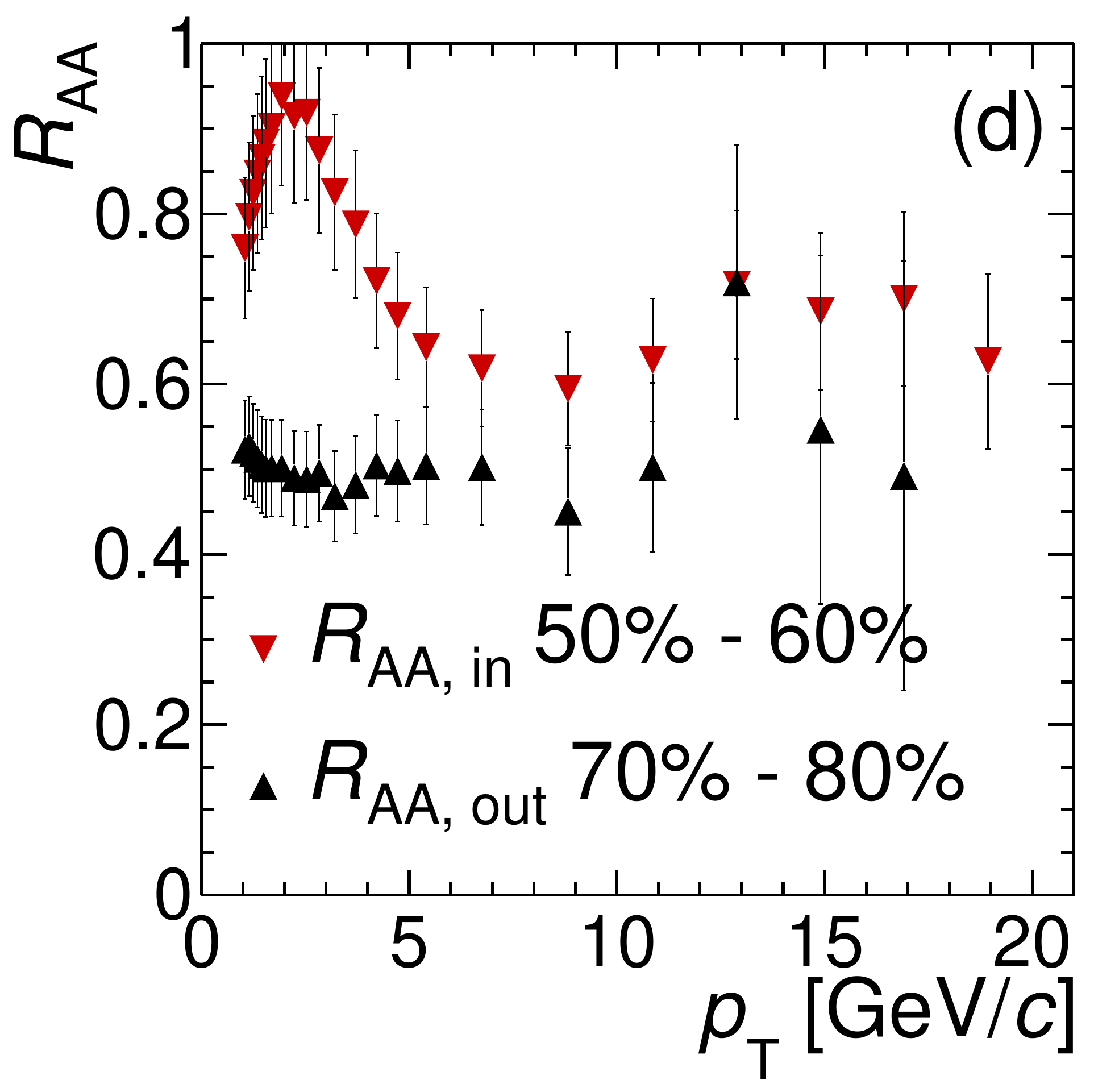}
  \end{center}
  \caption{(Color online) The comparison between \RAA in- and out-of-plane
    for situations where the scaling variable $\rhoc^\onehalf L$ is approximately the
    same. As can be seen, the good agreement observed in
    Fig.~\ref{fig:scaling} is reproduced at higher \pT.}
  \label{fig:pt_ratio}
\end{figure}

Furthermore, in Fig.~\ref{fig:pt_ratio} we demonstrate that the proposed
scaling variable, $\rhoc^\onehalf L$, seems to work reasonably well for all
\pT. Whereas the agreement is good for central collisions, one observes some
tension for the 70--80~\% centrality class. In the most peripheral collisions
it is known that the difference between the reaction plane and the impact
parameter plane is the largest so that one is more sensitive to the
description of individual collisions in the model. The impact of hard
scatterings on the experimental measurement of \vtwo could also be
significant due to the smaller number of participants.

\begin{figure}[htbp]
  \begin{center}
    \includegraphics[keepaspectratio, width=0.98\columnwidth]{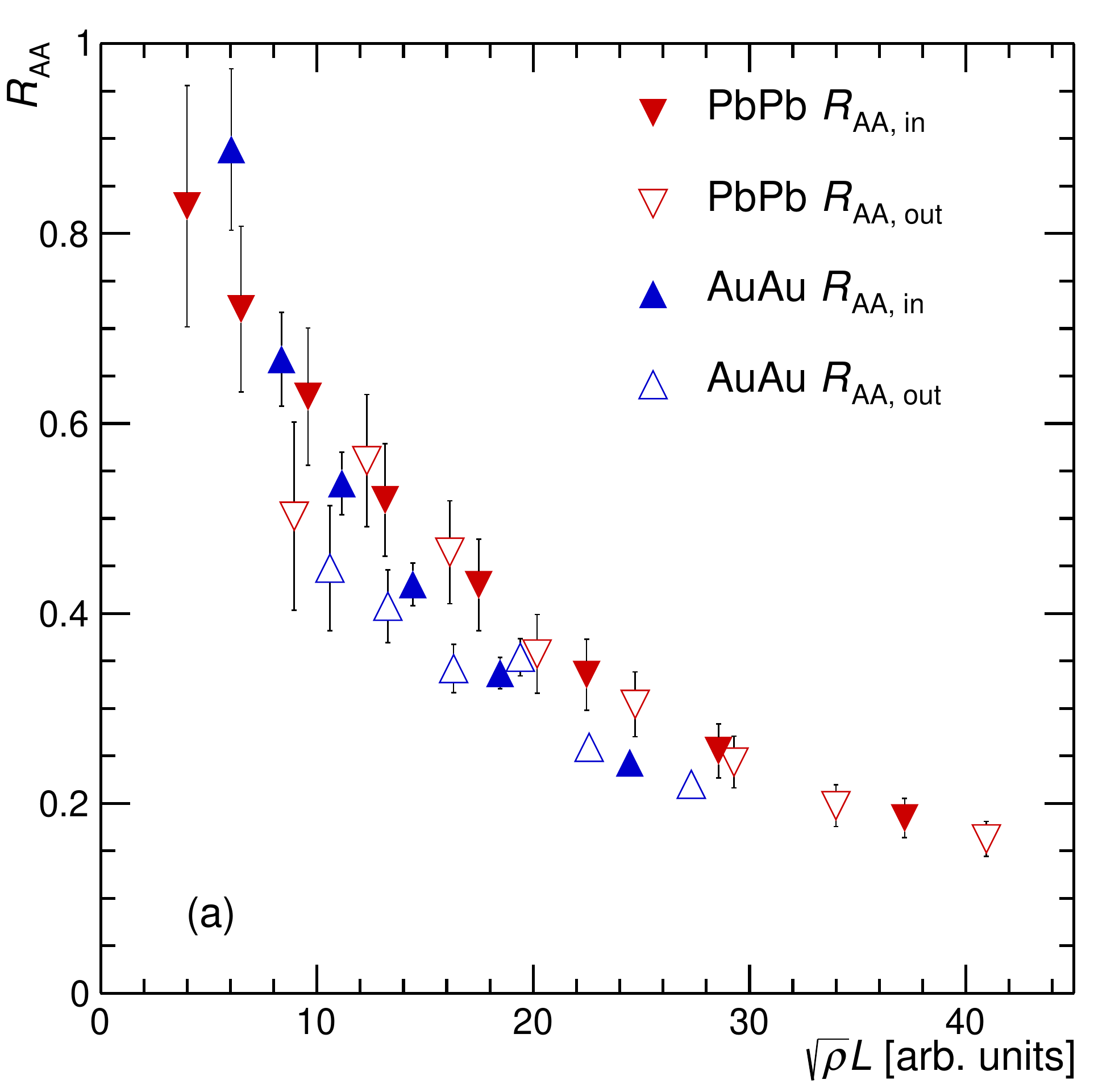}
    \includegraphics[keepaspectratio, width=0.98\columnwidth]{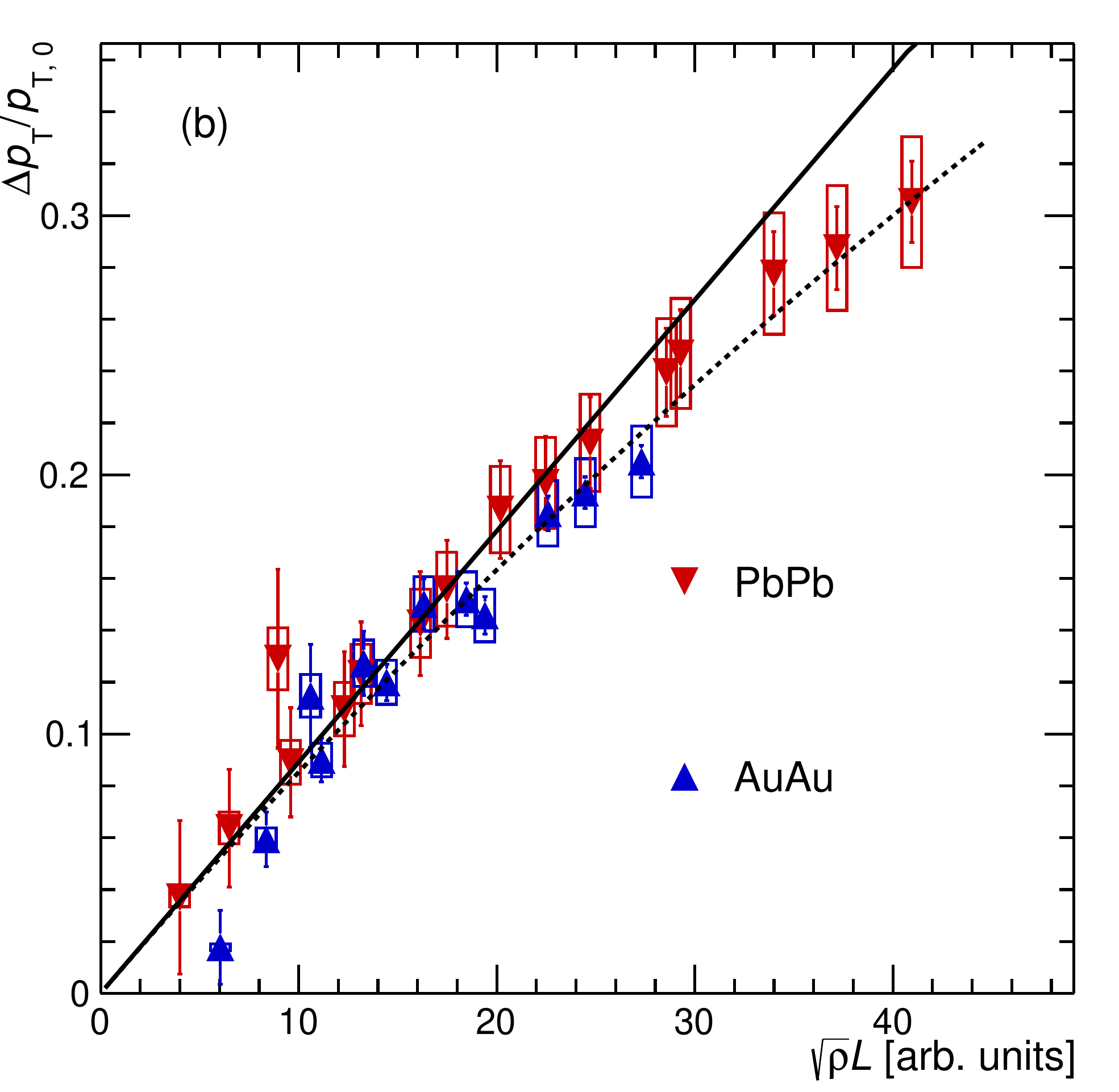}
  \end{center}
  \caption{(Color online) \RAA in- and out-of-plane for $\pT \sim \gevc{10-13}$ at \snnt{2.76}
    (red points) and for $\pT > \gevc{10}$ at \snng{200} (blue points) as a
    function of $\rhoc^\onehalf L$ (a). The corresponding \pT shifts as a
    function of the same scaling variable are shown in (b). Due to the different shape of the
    \pp spectrum the energy loss is the same in our model even if the \RAA is
    different.}
  \label{fig:finalResults}
\end{figure}

In the remainder of this section we will show that the scaling variable found above also works
surprisingly well both at RHIC and LHC. Recently PHENIX has
published the \RAA vs. the event plane at very high \pT for
\pizero~\cite{Adare:2012wg}. One should note that \pT spectra in \pp at RHIC
and LHC are power law-like for $\pT > \gevc{5}$, but that the power law
exponent is quite different in the two cases. The relationship between \RAA and
energy loss at LHC and RHIC is therefore different even if the \RAA are quite
compatible for each of the centralities. The main change in the scaling variable going
from LHC down to RHIC energies is an almost centrality independent decrease of particle density
$dN/d\eta$ of a factor 0.48~\cite{Aamodt:2010cz}. In our picture one therefore
expects the energy loss to be approximately 40\% larger at LHC than at RHIC
for similar centralities. This is very similar to what was found
in~\cite{Adare:2012wg}.
Figure~\ref{fig:finalResults} demonstrates that while the \RAA as a function of the
proposed scaling variable, $\rhoc^\onehalf L$, is different at LHC and RHIC, see the left panel,
the derived energy losses (which takes into account the difference in the
power law exponents) fall on a single curve as a function of the scaling
variable, see the right panel. We have fitted the \pT shift using two
parameterizations.\footnote{The two fits are a linear, $\DpT/\pT = C \xi$, and
  a non-linear relation, found by solving $\dd\pT/\pT = C \dd\xi$, where $\xi
  = \rhoc^\onehalf L$ and $C$ is the slope parameter. The latter parameterization illustrates that the deviation from the linear dependence on the scaling variable $\xi$ is consistent with a constant relative energy loss.} 
The deviation from a linear relation is only modest.

\section{Discussion}
\label{sec:discussion}

In the sections above, we have extracted a quite robust scaling law relating
the characteristic \pT shift of high \pT hadronic spectra in \nuclnucl
collisions to generic properties of the collision, such as the multiplicity
density and the RMS of its distribution, that seems to work over an order of
magnitude in collision energy. Despite the fact that these properties are
quite inclusive and do not take account of the dynamical evolution of the
system created in these collisions, the observed scaling suggests a dominant
and consistent mechanism underlying the physics of jet quenching from RHIC to
LHC.

In the discussion of energy loss we have focused on the very high \pT data while in Fig.~\ref{fig:pt_ratio} one clearly observes large differences at lower \pT ($\lesssim \gevc{6}$). Some of those can be attributed to the typically much larger flow in-plane than out-of-plane. It is important to note that the good agreement at high \pT shows that the density variation seems to be pivotal for the quenching mechanism, see Fig.~\ref{fig:scaling}. This might suggest that the transverse expansion of the medium has little effect on jet quenching, i.e., the dilution of the medium is canceled by the longer path length. This important issue certainly deserves further studies.

It is tempting to interpret the results from Sec.~\ref{sec:results} in light of radiative energy loss, see Appendix~\ref{sec:app:radiative} for a brief review. Note firstly that the na\"ive identification of the \pT shift with the mean energy loss taken by one-gluon emission, which would lead to $\DpT \sim \hat q L^2 \sim \rhoc^\threequarters L^2$, cf. Eq.~(\ref{eq:Shard}), fails to produce a scaling, see the bottom-right panel of Fig.~\ref{fig:scaling}. Accounting for multi-gluon emissions and the bias due to the steeply falling parton spectrum one rather expects $\DpT \sim \rhoc^\threeeights L$, cf. Eq.~(\ref{eq:Stypical}), which is close to what we observe in the data.\footnote{To study the expected \pT behavior of the shift from radiative processes, $\DpT \sim \pT^\onehalf$ goes beyond the scope of our present study.}

Similar studies have, as mentioned before, been carried out by Lacey \emph{et al.}~\cite{Lacey:2009ps,Lacey:2012bg,Lacey:2012kb}. The main difference from our work is that in their studies they do not take the density effect for different centralities into account and they obtain a single curve for \RAA vs. path length $L$. But, as can be seen in the top-left panel of Fig.~\ref{fig:scaling}, this relation breaks down when one studies \RAA in- and out-of-plane. Therefore their results should be supplemented by the additional information we have extracted here. There are also important differences in the physical pictures one extracts. Based on their findings they assert that jet quenching first sets in after a time of $\approx 1~\text{fm/$c$}$~\cite{Lacey:2009ps}. In our analysis, the intercept in the right panel of Fig.~\ref{fig:finalResults} is consistent with zero suggesting that the plasma formation time does not play a role for quenching. 

\begin{figure}[tbp]
  \begin{center}
    \includegraphics[keepaspectratio, width=0.99\columnwidth]{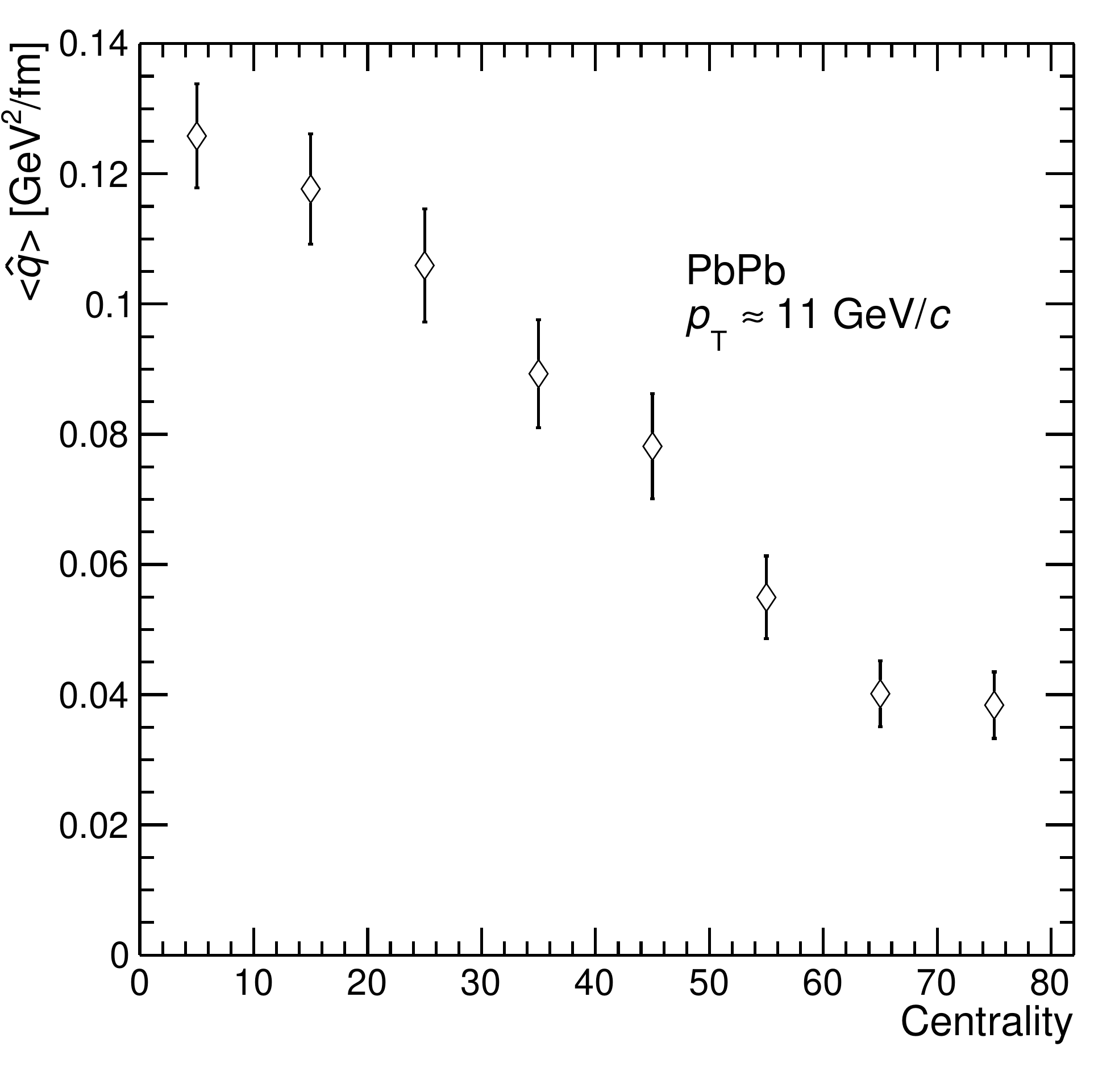}
  \end{center}
  \caption{ The $\langle \hat q \rangle$ as a function of centrality for LHC
    data extracted using Eq.~\ref{eq:qhat}.}
   \label{fig:qhat}
\end{figure}

We point out that our improved data driven analysis also allows to extract some information about the centrality dependence of the quenching phenomenon. Presently, we will identify the extracted density \rhoc from Eq.~(\ref{eq:BjorkenDensity}) with the transport parameter for jet quenching averaged over the trajectory of the jets, $\langle \hat q\rangle$, in the context of radiative energy loss. Then, from Eq.~\ref{eq:Stypical}, we find
\begin{equation}
\label{eq:qhat}
\langle \hat q \rangle = \left ( \frac{1}{L}\frac{\DpT}{\pT} \right )^{2}
\frac{n \pT}{4 \pi \bar \alpha^2} \,,
\end{equation}
where $n$ is the power of the invariant \pp spectrum and $\bar \alpha= \alpha_s C_R/\pi$ ($C_R$ being the relevant color factor), and we refer to Appendix~\ref{sec:app:radiative} for further details.
Figure~\ref{fig:qhat} displays the resulting centrality behavior, with $\bar \alpha = 0.3$
  and $\pT = \gevc{11}$. However, we note that this interpretation of the
  data driven results introduces some conceptual issues. 
In fact, we expect both $n$ and $\bar \alpha$ to vary with the center-of-mass collision energy. The reason for the variation of the latter quantity, comes about since at RHIC (LHC) we
  expect the high \pT particles to be fragments from dominantly quarks
  (gluons) implying a different color factor in $\bar \alpha$. 
  The similarity between RHIC and LHC in Fig.~\ref{fig:finalResults} therefore appears accidental in this context.
  We recall that the main motivation behind the data driven study was to avoid these conceptual difficulties.
  In our opinion, the most solid conclusion that can be drawn
  from Fig.~\ref{fig:qhat} is the decrease of $\langle \hat q \rangle$ by
  roughly a factor 4 from central to peripheral collisions dictated by the
  $\sqrt{\rhoc}$ dependence.

Albeit the data-driven analysis and subsequent interpretation both deal with
static quantities, and therefore are inherently consistent, a serious caveat
of the interpretation in terms of radiative energy loss is the neglecting of
the longitudinal expansion of the medium.
This can be estimated by making use of the dynamical scaling law for
$\hat q$ \cite{Salgado:2002cd,Salgado:2003gb}. For a Bjorken-expanding medium
the average transport parameter $\langle \hat q \rangle$ is related to the
initial $\hat q_0$ measured at some initial proper time $\tau_0$ as
$\langle\hat q \rangle \sim \tau_0 \hat q_0/L$.
This, in turn, implies that the expected path length dependence due to medium-induced radiative processes would scale as $\sim L^\onehalf$, rendering it incompatible with the extracted scaling behavior.
Within our data driven approach, these ideas rather imply that the extracted values of the average transport parameter involves a significantly larger
initial $\hat q_0$ in the early stages of the collision.
A generic theory driven approach
to a wide array of energy loss scenarios were presented in \cite{Betz:2012qq}
in the context of a Monte-Carlo model which also includes realistic nuclear
geometry and couples to a hydrodynamical model of the plasma, see also, e.g.,
\cite{Armesto:2009zi} for similar efforts.

The extraction of the \pT loss is done for charged particles while the
quenching supposedly affects the spectra at the parton level. The charged
particle \pT spectrum at high \pT largely reflects leading particles and as we
know from measurements at LHC that leading particle fragments in quenched and
unquenched jets share similar fractions of the jet
\pT~\cite{Chatrchyan:2012gw}, this approximation is probably not so bad. Still
it would be interesting to make a similar study with jets.

\section{Conclusions}
\label{sec:conclusions}
\raggedbottom

In this study the goal have been to distance ourselves as far as possible from
models of jet quenching and rather by selecting samples from different
centrality classes with similar path lengths to be able to isolate the density
effect and then study the path length dependence. Surprisingly the method
works very well and is in fact in reasonable agreement with theoretical
considerations. A critical question is how the longitudinal expansion of the
medium affects jet quenching and this has tremendous impact on how one would
interpret the results in terms of e.g. the path length dependence. 

Finally we note that the exact same density dependence observed for different
centrality classes for LHC data is consistent with RHIC data indicating that
the dense matter at RHIC and LHC has fundamentally similar properties.

\acknowledgements{

PC wishes to express his gratitude to the Swedish Research Council
  for financial support. KT is supported by a Juan de la Cierva fellowship
and by the research grants FPA2010-20807, 2009SGR502, the Consolider
CPAN project and FEDER.  }

\appendix
\section{How to estimate the \pT shift}
\label{sec:app:ptloss}

The definition of the nuclear modification factor is
\begin{equation}
\RAA(\pT) = \frac{\dd N_\text{AA} \big/ \dd \pT}{N_\text{coll} \, \dd N_\text{pp} \big/ \dd \pT } \,,
\end{equation}
where $N_\text{coll}$ represents the number of binary collisions (the nuclear overlap function) for the given centrality class estimated from the Glauber model (see, e.g., \cite{Miller:2007ri}).
Following the standard interpretation of the suppression of hadron spectra in \nuclnucl collisions, we assume that it arises due to a \pT shift of the primordial parton spectrum. We will therefore write
\begin{equation}
\label{eq:AAspectrum}
\frac{\dd N_\text{AA}(\pT)}{\dd \pT} = N_\text{coll} \frac{\dd N_\text{pp} \left(\pT' = \pT + \dpT \right)}{\dd \pT'} \,\left|\frac{\dd \pT'}{\dd \pT}\right| \,,
\end{equation}
where we have made explicit for which \pT value the spectrum is evaluated at and included the Jacobian of the transformation, which also can be written as $\dd \pT' \big/ \dd \pT = 1 + \dd \dpT \big/ \dd \pT$. Thus, the Jacobian differs from unity if \dpT is a function of \pT. Explicitly, the spectrum on the LHS of Eq.~(\ref{eq:AAspectrum}) is measured at a given \pT, while $\pT'$ on the RHS represents the primordial momentum of the parton prior to energy loss. Thus, the master equation to extract the energy loss via the \pT shift reads
\begin{equation}
\frac{\dd N_\text{pp}(\pT')}{\dd \pT'} = \RAA(\pT) \frac{\dd N_\text{pp}(\pT)}{\dd \pT}\,\left|\frac{\dd \pT}{\dd \pT'}\right|
\end{equation}
Having no \emph{a priori} knowledge about the specific form of \dpT that enters the Jacobian, we will parameterize it using two ``extreme" cases:
\begin{enumerate}
\item Firstly, we assume that $\pT = k \, \pT'$, where $0<k<1$ is a constant. This implies that
\begin{equation}
\pT'\frac{\dd N_\text{pp}(\pT')}{\dd \pT'} = \RAA(\pT) \,\pT\frac{\dd N_\text{pp}(\pT)}{\dd \pT} \,.
\end{equation}
\item Secondly, we assume a constant \pT shift, $\dpT = \text{const.}$ The Jacobian is simply unity, and we get that
\begin{equation}
\frac{\dd N_\text{pp}(\pT')}{\dd \pT'} = \RAA(\pT) \,\frac{\dd N_\text{pp}(\pT)}{\dd \pT} \,.
\end{equation}
\end{enumerate}
Relevant cases, for which typically $\dpT \sim \pT^\alpha$ where $0<\alpha<1$
(e.g., see Eqs.~(\ref{eq:Stypical}) and (\ref{eq:Shard})), fall in between the
``extremes" considered above. The \pT shifts estimated from these two cases
will be averaged and the difference will be indicated as a systematic
  uncertainty of the procedure.

\section{Radiative energy loss}
\label{sec:app:radiative}

For highly energetic probes the hot and dense medium is parameterized by one characteristic transport coefficient, the so-called $\hat q$ parameter which encodes the transverse momentum broadening per unit length. Heuristically, this parameter scales with the energy density $\rho$ as $\hat q \propto \rho^\threequarters$.
The largest energy that can be carried by a medium-induced gluon accumulates momentum along the whole path length of the medium
and is usually defined as $\omega_c \equiv \hat q L^2/2$. The spectrum of induced gluons per unit length reads \cite{Zakharov:1997uu,Baier:1998kq}
\begin{equation}
\label{eq:BDMPSspectrum}
\omega \frac{\dd I}{\dd \omega \,\dd L} = \bar \alpha \sqrt{\frac{\hat q}{\omega}} \,,
\end{equation}
for energies $\omega < \omega_c$,\footnote{To be precise, the spectrum in Eq.~(\ref{eq:BDMPSspectrum}) is regularized at a minimal energy marking the onset of the Bethe-Heitler regime.} where $\bar \alpha \equiv \alpha_s C_R/\pi$. It follows that the energy loss caused by the single-gluon emission, given by $
-\dd E\big/\dd L = \bar \alpha \hat q L$, is dominated by the hard sector, $\omega \sim \omega_c$. One should on the other hand keep in mind that the number of gluons, given by $N(\omega) \sim \sqrt{\bar \alpha^2 \omega_c/\omega}$, becomes large for soft gluons, in particluar, when $\omega < \bar \alpha^2 \omega_c$.

The quenching factor, which encodes the partonic spectrum modified in the medium prior to fragmentation,\footnote{See \cite{CasalderreySolana:2012ef} for a discussion of the validity of such an assumption. For our present purposes, the quenching factor serves as a good indicator of the parametric behavior of the nuclear modification factor \RAA.} is defined as
\begin{equation}
Q(\pT) \equiv \int_0^\infty \dd \epsilon \,D(\epsilon) \,\frac{\dd^2 \sigma^\text{vac}(\pT + \epsilon)\big/{\dd \pT}^2}{\dd^2 \sigma^\text{vac}(\pT)\big/{\dd \pT}^2} \,,
\end{equation}
where $D(\epsilon)$ is the probability distribution of energy loss.  Assuming independent gluon emissions it is simply given by a Poisson distribution \cite{Baier:2001yt}, but this premise can be improved upon by including, e.g., phase-space limitations \cite{Salgado:2003gb} or energy-momentum conservation, see \cite{Blaizot:2012fh,Blaizot:2013hx}. These corrected distributions give rise to more complex scaling trends than discussed below, but will be neglected in the following. Presently we assume that the invariant \pp is well described by a power law spectrum with constant exponent $n$. Then, in the large-$n$ approximation we recast the quenching factor as $Q(\pT) = \exp (- n \dpT \big/\pT)$,
where \dpT is directly related to the \pT shift of the medium-modified parton spectrum as $\dd^2 \sigma^\text{med}(\pT) \big/{\dd \pT}^2 = \dd^2 \sigma^\text{vac}(\pT + \dpT) \big/{\dd \pT}^2$. This shift can be estimated to be \cite{Baier:2001yt}
\begin{equation}
\label{eq:Stypical}
\dpT =\int_0^\infty \dd \omega \,N(\omega) \exp \left( -\frac{n \omega}{\pT} \right) \approx \sqrt{\frac{8 \pi\, \bar \alpha^2 \,\omega_c \,\pT}{n}} \,.
\end{equation}
Inserting the latter expression into the formula for $Q(\pT)$ we obtain the so-called ``pocket formula" for radiative energy loss \cite{Baier:2001yt,Dokshitzer:2001zm,Lacey:2009ps,Lacey:2012bg,Lacey:2012kb}. Relating to our previous discussion, the shift scales as $\dpT \sim \pT^\onehalf \rho^\onehalf L$.

Finally, note that in the special limit of $\pT >n \omega_c$ the \pT shift rather becomes
\begin{equation}
\label{eq:Shard}
\dpT \simeq \int_0^\infty \dd \omega \,N(\omega) \sim \omega_c \,,
\end{equation}
and scales as $\dpT \sim \rho^\threequarters L^2$. Thus, only in this particular regime can one identify the mean energy loss with the typical \pT shift due to the dominance of one-gluon emission.
The bias due to the steeply falling parton spectrum tend to shift the {\it typical} energy loss to smaller values, as given by Eq.~(\ref{eq:Stypical}).

\end{document}